\documentstyle[12pt]{article}

\def\endpage{\vfill\eject}

\newcommand{\AmS}{{\protect\the\textfont2
\renewcommand{\thesection}{\Roman{section}}
  A\kern-.1667em\lower.5ex\hbox{M}\kern-.125emS}}
 
\begin{document}
\rightline {\bf DFTUZ/96/4}
\vskip 2. truecm
\centerline{\bf The Continuum Limit of Non Compact QED}
\vskip 2 truecm
\centerline { V.~Azcoiti$^a$, G. Di Carlo$^b$, A. Galante$^{c,b}$, 
A.F. Grillo$^d$, V. Laliena$^a$ and C.E. Piedrafita$^b$}
\vskip 1 truecm
\centerline {\it $^a$ Departamento de F\'\i sica Te\'orica, Facultad 
de Ciencias, Universidad de Zaragoza,}
\centerline {\it 50009 Zaragoza (Spain).}
\vskip 0.15 truecm
\centerline {\it $^b$ Istituto Nazionale di Fisica Nucleare, 
Laboratori Nazionali di Frascati,}
\centerline {\it P.O.B. 13 - Frascati 00044 (Italy). }
\vskip 0.15 truecm
\centerline {\it $^c$ Dipartimento di Fisica dell'Universit\'a 
dell'Aquila, L'Aquila 67100 (Italy)}
\vskip 0.15 truecm
\centerline {\it $^d$ Istituto Nazionale di Fisica Nucleare, 
Laboratori Nazionali del Gran Sasso,}
\centerline {\it Assergi (L'Aquila) 67010 (Italy). }
\vskip 3 truecm
\centerline {ABSTRACT}
Since four-fermion operators in 
strongly coupled $QED$ are nonperturbatively renormalizable, we analyze here 
the phase diagram and critical behaviour of the Gauged Nambu-Jona Lasinio 
model. Our mean field approximation relates the critical exponents 
along the continuous phase transition line with the mass dependence of the 
chiral condensate in the Coulomb phase of standard noncompact $QED$. The 
numerical results for noncompact $QED$ strongly 
suggest non mean field exponents along the critical line.

\vfill\eject

\par
The possible existence of a non gaussian fixed point in strongly coupled 
non compact $QED$ has been subject of extensive research during the 
last few years. It is well known that it has been a controversial subject 
for a long time. The present status of the field can be summarized in the 
following two points:

1. Triviality as in a logarithmically improved scalar mean field theory 
has been disproved by the numerical results \cite{TODOS}. 

2. Unfortunately triviality a la Nambu-Jona Lasinio, which is the most 
natural way to analyze triviality in a theory with strongly coupled 
fermion fields, can not be disproved by the present day 
numerical data and it seems very difficult to get an important improvement 
in the quality of these data in near future.

Therefore other alternative ways and new ideas are necessary in order to 
get some progress in this field. This is the reason why the 
gauged Nambu-Jona Lasinio $(GNJL)$ model has become increasingly 
interesting in recent time. 
In fact if a non gaussian fixed point exists in non compact $QED$, 
the naive dimensional analysis does not applies. Therefore operators of 
dimension higher than four, which are non renormalizable in perturbation 
theory, could acquire anomalous dimensions and become renormalizable 
\cite{BARDEEN}. 

The lattice action for the $GNJL$ model with noncompact gauge fields and 
staggered fermions reads

$$S=\frac{\beta}{2}\sum_{n,\mu<\nu}\Theta_{\mu\nu}^2(n)\:+\:\bar\chi\Delta
(\theta)\chi\:+\:m\bar\chi\chi
\:-\:G\sum_{n,\mu}\bar\chi_n\chi_n\bar\chi_{n+\mu}\chi_{n+\mu}.
\eqno(1)$$

In the chiral limit, $m=0$, this action is invariant under the continuous 
transformations 

$$\chi_n\;\;\rightarrow\;\;\chi_n\: e^{i\alpha (-1)^{n_1+\ldots+n_d}}
\;\;\;\;\;\;\;
\;\;\bar\chi_n\;\;\rightarrow\;\;\bar\chi_n\: e^{i\alpha (-1)^{n_1+\ldots
+n_d}}
\eqno(2)$$

\noindent
which define a continuous chiral $U(1)$ symmetry group.

The main technical difficulty when computing vacuum averages in the 
$GNJL$ model comes from the fact that the action (1) is not a bilinear of 
the fermion fields. The standard procedure consists in the introduction of 
an auxiliary vector field which allows to bilinearize the fermion action. 
The prize to pay for that is that we have one more field to include in the 
numerical simulations of this model that besides the number of free parameters 
$(\beta, m, G)$, makes it difficult to analyse this model with reasonable 
computer resources \cite{EDIN}.

As a first approach, we can do a standard mean field approximation which has 
the advantage of bilinearizing the action (1). Following the mean field 
technique, we do in (1) the following substitution

$$G\sum_{n,\mu}\bar\chi_n\chi_n\bar\chi_{n+\mu}\chi_{n+\mu}\;\;\rightarrow
\;\;2dG\langle\,\bar\chi\chi\,\rangle
\sum_n\bar\chi_n\chi_n
\eqno(3)$$

\noindent
where $d$ is the space-time dimension. The action (1) becomes in this way a 
bilinear in the fermion fields and the path integral over the Grassmann 
variables can be done by mean of the Mathews-Salam formula.

The $v.e.v.$ of the chiral condensate after the substitution 
of the mean field approximation (3) in the 
action (1) is given by

$$\langle\,\bar\chi\chi\,\rangle
=-\frac{\int\,[d\theta]\,e^{-\frac{\beta}{2}\sum\Theta_{\mu\nu}^2(n)}
\det[\Delta+(m-8G\langle\,\bar\chi\chi\,\rangle)
I]\frac{1}{V}tr\frac{1}{\Delta+(m-8G\langle\,\bar\chi\chi\,\rangle)I}}
{\int\,[d\theta]\,e^{-\frac{\beta}{2}\sum\Theta_{\mu\nu}^2(n)}
\det[\Delta+(m-8G\langle\,\bar\chi\chi\,\rangle)I]},
\eqno(4)$$

In the chiral limit 
$m=0$, equation (4) becomes

$$\langle\,\bar\chi\chi\,\rangle=
16G\langle\,\bar\chi\chi\,\rangle\,
\left\langle\,\frac{1}{V}\sum_{j=1}^{V/2}\frac{1}
{\lambda_j^2+64G^2{\langle\,\bar\chi\chi\,\rangle}^2}\,\right\rangle
\eqno(5)$$

\noindent
where the sum in (5) runs over all positive eigenvalues of the massless 
Dirac operator and the integration measure in the $v.e.v.$ includes the 
fermionic determinant of standard noncompact $QED$, evaluated at the 
effective mass $\bar{m}=-8G\langle\,\bar\chi\chi\,\rangle$.

\vskip 1truecm
\leftline{\bf 1. The phase diagram}
\par
Equation (5) is always verified if $\langle\,\bar\chi\chi\,\rangle=0$, and 
this is the only solution in the symmetric phase. In the broken phase where 
$\langle\,\bar\chi\chi\,\rangle\neq 0$, the $v.e.v.$ of the chiral condensate 
will be given by the solution of the following equation

$$1=16G\,\left\langle\,\frac{1}{V}\sum_{j=1}^{V/2}\frac{1}
{\lambda_j^2+64G^2{\langle\,\bar\chi\chi\,\rangle}^2}\,
\right\rangle
\eqno(6)$$

\noindent
which gives for the critical line, where the chiral condensate vanishes 
continuously, the following expression

$$G_c(\beta)=\frac{1}{\frac{16}{V}\,\langle\,\sum_{j=1}^{V/2}\frac{1}
{\lambda_j^2}\,\rangle}
\eqno(7)$$

The existence of this critical line in the $GNJL$ model was discovered some 
time ago \cite{CRITICAL} in the continuum formulation using the 
quenched-ladder approximation.

In Fig. 1 we present our numerical results for the phase diagram in the 
$\beta, G$ plane. The critical line has been obtained by computing numerically 
the $v.e.v.$ of the sum of the inverse square eigenvalues (eq. (7)), which 
is proportional to the chiral transverse susceptibility of the standard 
noncompact $QED$ in the chiral limit. The numerical simulations where done 
using the $MFA$ approach \cite{MFA}, which 
allows to do computations in the chiral limit. We refer the interested 
reader to the extended bibliography on this subject \cite{MFA} 
and specially to the ref. \cite{SUSCEP} where the computation of the chiral 
susceptibility and the determination of the critical coupling in noncompact 
$QED$ is discussed in detail.

\vskip 1truecm
\leftline{\bf 2. The critical exponents}
\par
The phase diagram of Fig. 1 is in good qualitative agreement with the 
corresponding phase diagram obtained in the quenched-ladder 
approximation \cite{CRITICAL}. Using this analytical 
approach, a line of critical 
points with continuously varying critical exponents was found in \cite{HOMBRE}, 
the intersection point of this line with the $G=0$ axis corresponding to an 
essential singularity \cite{BARDEEN}.

Later on, numerical simulations of noncompact $QED$ disproved the essential 
singularity behavior, putting in evidence the limitations of the 
quenched-ladder approximation. Since our approach contains less 
approximations, we do hope to get more reliable results for the critical 
exponents.

In order to extract the critical exponents, we will start from the key 
equation of state relating the order parameter with the external 
magnetic field $m$ and the gauge and four fermion couplings. Using the 
previous notation we can write 

$$\langle\,\bar\chi\chi\,\rangle=-2\bar{m}F(\beta,\bar{m})
\eqno(8)$$

\noindent
where the right hand side in (8) is just the chiral condensate in full 
noncompact $QED$ evaluated at the gauge coupling value $\beta$ and fermion 
mass $\bar{m}$. Concerning critical exponents the interesting physical 
region, as follows from the phase diagram of Fig. 1, is $\beta>\beta^{0}_c$ 
(Coulomb phase of noncompact $QED$).

Since we are interested in the critical region ($m\rightarrow 0$, 
$\langle\,\bar\chi\chi\,\rangle\rightarrow 0$), we will analyze the behavior 
of $F(\beta,\bar{m})$ in the $\bar{m}\rightarrow 0$ limit. In this limit 
we can write

$$F(\beta,\bar{m})=F(\beta,0) + B\bar{m}^\omega\:+\:\ldots
\eqno(9)$$

The second term in (9) can contain also logarithmic 
contributions and $F(\beta,0)$ is one half of the massless transverse 
susceptibility in noncompact $QED$.
Equation (9), after the substitution of $\bar{m}$ by 
$m-8G\langle\,\bar\chi\chi\,\rangle$, implies the following behavior for 
the chiral condensate in the $m\rightarrow 0$ limit

$$\langle\,\bar\chi\chi\,\rangle\sim m^{\frac{1}{\omega+1}}
\eqno(10)$$

\noindent
and therefore the $\omega$ and $\delta$ exponents are related by the equation 

$$\delta = \omega + 1
\eqno(11)$$

A straightforward calculation allow to compute also the 
magnetic $\beta_m$ and susceptibility $\gamma$ exponents, the final result 
being

$$\beta_m=\frac{1}{\omega},\;\;\;\;\;\;\;\gamma = 1
\eqno(12)$$

The hyperscaling relation $\gamma=\beta_{m}(\delta-1)$ is verified, as follows 
from (12).

The determination of the order parameter critical exponents in our mean field 
approach reduce therefore to the determination of the $\omega$ exponent which 
controls the mass dependence of the chiral condensate in the Coulomb phase 
of noncompact $QED$. In the $\beta\rightarrow\infty$ limit of noncompact 
$QED$, the theory is free and the chiral condensate can be analytically 
computed. The well known result in this case ($\omega=2$ plus logarithmic 
corrections) implies mean field exponents for the end point of the phase 
transition line, with the following behavior for 
$\langle\,\bar\chi\chi\,\rangle_{\beta=\infty,G=G^{\infty}_c}$

$$m\sim {\langle\,\bar\chi\chi\,\rangle}^{3} 
\log \langle\,\bar\chi\chi\,\rangle
\eqno(13)$$

In the general case, the chiral condensate in the Coulomb phase of 
noncompact $QED$ $(\langle\,\bar\chi\chi\,\rangle_{NCQED})$ can be 
parameterized as follows

$$\langle\,\bar\chi\chi\,\rangle_{NCQED} = A(\beta) m + B(\beta) m^
{\omega+1}\:+\:\ldots
\eqno(14)$$

The first contribution in (14) is linear in $m$, as follows from the fact 
that the massless transverse susceptibility is finite in the Coulomb phase 
of noncompact $QED$. The next contribution can have logarithmic 
corrections, as happens in the $\beta\rightarrow\infty$ limit where it becomes 
$m^{3}\log m$. In order to extract the $\omega$ exponent from the numerical 
simulations, we can use the results for the massless chiral transverse 
susceptibility \cite{SUSCEP} 
to fix $A(\beta)$ in (14) and fit the numerical results with 
eq. (14). This procedure has the inconvenient that higher order contributions 
in (14) can induce systematic errors in the determination of $\omega$.

A better strategy exploiting the potentialities of the $MFA$ method is 
to compute vacuum 
expectation values of operators which can be considered as 
generalizations of one of the contributions to the massless nonlinear 
susceptibility. More precisely 
we have defined $\chi_q$ by the expression

$$\chi_{q} = \frac{1}{V}\left\langle\,\sum_{j=1}^{V/2}\frac{1}
{\lambda_j^q}\,\right\rangle
\eqno(15)$$

When $q=4$, we get one of the contributions to the standard massless 
nonlinear susceptibility. In the 
general case we can write this vacuum expectation value as an integral 
over the spectral density of eigenvalues in the following way

$$\frac{1}{V}\left\langle\,\sum_{j=1}^{V/2}\frac{1}
{\lambda_j^q}\,\right\rangle = \int \frac{\rho(\lambda)}{\lambda^q} 
d\lambda
\eqno(16)$$

\noindent
and if the density of eigenvalues $\rho(\lambda)$ behaves like $\lambda^p$ 
near the origin, $\chi_q$ will diverge when $q>p+1$. In such a case and 
for lattices of finite size, we expect for $\chi_q$ the following type 
of divergency with the lattice size $L$

$$\chi_{q} \sim L^{\alpha(q-p-1)}
\eqno(17)$$

\noindent
where $\alpha$ in (17) is some positive number.

An interesting thing to notice is that the p-exponent which controls the 
small $\lambda$ behavior of the spectral density $\rho(\lambda)$, can be 
related with the $\omega$ exponent by the following equations 

$$\omega = p-1 (p\leq 3)$$

$$\omega = 2 (p>3)
\eqno(18)$$

These relations allow to extract the $\omega$ exponent from the finite size 
behavior of the generalized nonlinear susceptibility $\chi_q$.

In Fig. 2 we have plotted our results for the inverse of the generalized 
nonlinear susceptibility $\chi_q$ against the inverse lattice size 
for q-values running from 2 to 4 and $\beta=0.237$. The solid lines in this 
figure correspond to a fit of all the points at any fixed q with the function 

$$\chi_{q}^{-1}(L) = a_q+b_q L^{-c}
\eqno(19)$$ 

The results reported in this figure show how the infinite volume limit of 
the inverse generalized nonlinear susceptibility vanishes at large q and 
is different from zero at small q, as expected. Fig. 3 is a plot of the 
extrapolated values of $\chi_q^{-1}$ (thermodynamical limit) against q. The 
critical value of q at which $\chi_q^{-1}$ vanishes can be estimated from 
these results. Hence we get $q_c\sim 2.5$ at $\beta=0.237$, which implies 
$p\sim 1.5$ and $\omega\sim 0.5$. Using now the relations (11), (12) the 
following results for the order parameter critical exponents can be derived

$$\delta\sim 1.5,\;\;\;\;\;\;\;\beta_{m}\sim 2,\;\;\;\;\;\;\;\gamma=1, 
\eqno(20)$$

\noindent
values which are clearly outside the range of the mean field exponents. 

Even if at first sight non mean field exponents in a mean field 
approach seems to be a paradox, there is no contradiction between both 
statements. In fact we have applied the mean field approximation to the 
fermion field but fluctuations of the gauge field are taken into account 
in our numerical simulations. In the infinite $\beta$ limit, where the 
gauge field is frozen to the free field configuration, we get mean field 
exponents. However fluctuations of the gauge field at finite $\beta$ seem 
to play a fundamental role in driving critical exponents to non mean field 
values.

The picture which emerges from this calculation is that the critical exponents 
change continuously along the critical line of Fig. 1 from their mean 
field values (end point of the critical line) to some non mean field values 
at the critical point of noncompact $QED$. The $\delta$ exponent approaches 
its mean field value $(\delta=3)$ from below whereas the magnetic exponent 
approaches its mean field value $(\beta_{m}=0.5)$ from above \cite{SACHA}. 
Our results for several values of the gauge coupling $\beta$ 
suggest also that the value of $\delta$ increases systematically 
along the critical line with increasing $\beta$, in contrast with the magnetic 
exponent results which are systematically decreasing with $\beta$. 

In conclusion we do believe our qualitative picture is realistic. In fact it is 
hard to think that non mean field exponents in a mean field approach will 
become mean field exponents after removing the mean field approach.

\endpage

\endpage
\vskip 1 truecm
\leftline{\bf Figure captions}
\vskip 1 truecm
{\bf Figure 1.} Phase diagram of the $GNJL$ model in the 
$\beta, G$ plane.

{\bf Figure 2.} Logarithm of the nonlinear susceptibility 
against the logarithm of the lattice size for lattice sizes 4,6,8, 10 and 
$\beta=0.237$.

{\bf Figure 3.} Inverse generalized nonlinear 
susceptibility against the inverse lattice volume at $\beta=0.237$.

{\bf Figure 4.} Infinite volume limit of the generalized 
nonlinear susceptibility against q at $\beta=0.237$.


\vskip 1 truecm
\begin{thebibliography}{9}
\bibitem{TODOS}
A. Kocic, J.B. Kogut and K.C. Wang, 
Nucl. Phys. {\bf B398} \rm (1993) 405;
A.M. Horowitz, 
Phys. Lett. {\bf 244B} \rm (1990) 306;
M. G\"okeler, R. Horsley, P. Rakow, G. Schierholz and R. Sommer, 
Nucl. Phys. {\bf B371} \rm (1992) 713;
V. Azcoiti, G. Di Carlo and A.F.Grillo,
Int. J. Mod. Phys. {\bf  A8} \rm (1993) 4235. 
\bibitem{BARDEEN}
P.I. Fomin, V.P. Gusynin, V.A. Miransky and Yu. A. Sitenko, 
Riv. Nuovo Cim. {\bf 6} \rm (1983) 1; 
V.A. Miransky, 
Nuovo Cim. {\bf 90A} \rm (1985) 149; 
C.N. Leung, S.T. Love and W.A. Bardeen, 
Nucl. Phys. {\bf B273} \rm (1986) 649.
\bibitem{EDIN}
S.P. Booth, R.D. Kenway, B.J. Pendleton, 
Phys. Lett. {\bf B228} \rm (1989) 115.
\bibitem{CRITICAL}
K.I. Kondo, H. Mino, K. Yamawaki, Phys. Rev. {\bf D39} \rm (1989) 2430;
T. Appelquist, M. Soldate, T. Takeuchi, L.C.R. Wijewardhana, in 
{\em Proceedings of John Hopkins Workshop on Current Problems in Particle 
Theory}, eds. G. Domokos and S. Kovesi-Domokos. (World Scientific 
Publishing Co. Singapore 1988).
\bibitem{MFA}
V. Azcoiti, G. Di Carlo and A.F. Grillo, 
Phys. Rev. Lett. {\bf 65} \rm (1990) 2239;
V. Azcoiti, A. Cruz, G. Di Carlo, A.F. Grillo and A. Vladikas, 
Phys. Rev. {\bf D43} \rm (1991) 3487; 
V. Azcoiti, G. Di Carlo, L.A. Fernandez, A. Galante, A.F. Grillo, V. Laliena, 
X.Q. Luo, C.E. Piedrafita and A. Vladikas; 
Phys. Rev. {\bf D48} \rm (1993) 402. 
\bibitem{SUSCEP}
V. Azcoiti, G. Di Carlo, A. Galante, A.F. Grillo, V. Laliena and 
C.E. Piedrafita, 
"Chiral susceptibilities in non compact $QED$: a new determination of the 
$\gamma$ exponent and critical couplings"; 
{\bf DFTUZ/95/12} \rm (1995). 
\bibitem{HOMBRE}
A. Kocic, S. Hands, J.B. Kogut, E. Dagotto, 
Nucl. Phys. {\bf B347} \rm (1990) 217.
\bibitem{SACHA}
A. Kocic, 
Nucl. Phys. {\bf B34} (Proc. Suppl.) \rm (1994) 129; 
A. Kocic, J.B. Kogut, 
Nucl. Phys. {\bf B422} \rm (1994) 593
\end{thebibliography}
\end{document}